\documentstyle[nato,numreferences,epsf,graphicx]{crckapb}

\newcommand{\chibar}{{\overline{\chi}}}
\newcommand{\psibar}{{\overline{\psi}}}
\newcommand{\Dsl}{D\!\!\!\!/}
\newcommand{\dsl}{\partial\!\!\!/}
\newcommand{\tk}{\tilde\kappa}
\newcommand{\tr}{\tilde r}

\begin{opening}
\title{Lattice chiral gauge theories through gauge fixing}

\author{Maarten Golterman}
\institute{Department of Physics and Astronomy\\
           San Francisco State University\\
           1600 Holloway Ave, San Francisco, CA 94132, USA}

\author{Yigal Shamir}
\institute{School of Physics and Astronomy\\
           Tel-Aviv University, Ramat Aviv 69978, Israel}

\end{opening}

\begin{document}

\begin{abstract}
After an introduction in which we review the fundamental
difficulty in constructing lattice chiral gauge theories,
we discuss the analytic and numerical evidence that abelian
lattice chiral gauge theories can be non-perturbatively constructed
through the gauge-fixing approach.  While a complete
non-abelian extension is still under construction, we
also show how fermion-number violating processes are realized in
this approach.\footnote{The first part overlaps significantly
with part of ref.~\cite{dubna}; the latter part is new.}
\end{abstract}

\vskip -0.5cm
\section{Introduction}

We start reviewing the difficulties
underlying the construction of lattice chiral gauge theories
(ChGTs).
These difficulties go back to the fundamental
observation by Nielsen and Ninomiya known as the ``species
doubling" theorem \cite{nn}, and by Karsten and Smit on the
role of the chiral anomaly \cite{ks}.

Consider a collection of left-handed fermion fields transforming
in a
representation of some symmetry group.\footnote{In four dimensions,
we can always take all fermions left-handed.}
A gauge theory containing these fermions can be regulated by putting it on a
lattice.  We may then investigate the anomaly structure
of the theory by keeping the gauge fields external (and smooth).
It is clear that each fermion field will have to contribute
its share to the expected chiral anomaly.  This can happen in
two ways: either the regulated theory is exactly invariant under
the symmetry group, and each fermion comes with its species
doublers, or the symmetry is explicitly broken
by the regulator ({\it i.e.} the lattice), making it possible
for each fermion field
to produce the correct contribution to the anomaly in the continuum
limit ({\it i.e.} for smooth gauge fields).

The Nielsen-Ninomiya theorem tells us that fermion
representations with doublers contain equally many left- and right-handed (LH and RH)
fermions transforming the same way under the symmetry group.
This way, the theory is anomaly free, and the doublers thus provide
the mechanism through which the symmetry group can remain an
exact invariance on the lattice.  The price one pays, however, is
that if we now make the gauge fields dynamical, a vector-like
gauge theory will emerge.

This means that, if we wish to construct a genuinely
chiral theory on the lattice, we have two options.  Either we
modify the symmetry group on the lattice so as to ``circumvent"
the Nielsen--Ninomiya theorem, or we introduce an explicit breaking
of the symmetry group.

The first option leads to a discretization of the Dirac
operator that satisfies the Ginsparg--Wilson
relation \cite{gw}, and will not be the subject of this talk.
For a general review, including an argument as to
how the modification of chiral symmetry on the lattice leads
to the Ginsparg--Wilson relation, see ref.~\cite{mg00}.
A proposal on how to apply these ideas toward a construction
of lattice ChGT is reviewed in
refs.~\cite{ml99,mg00}.  While a non-perturbative construction
for abelian chiral theories based on this approach exists,
it is still an open question whether it can also be generalized
to the non-abelian case.

A well-known example of the second option is the
formulation of lattice QCD with Wilson fermions \cite{wilson}.
In this method, a momentum-dependent Wilson mass term of the form
\begin{equation}
-\frac{r}{2}\sum_{x\mu}\left(\psibar_x\psi_{x+\mu}+\psibar_{x+\mu}\psi_{x}
-2\psibar_x\psi_x\right)  \label{WMASS}
\end{equation}
is added to the action, which removes the doublers by giving them a
mass of order $1/a$ (where $a$ is the lattice spacing,
taken equal to one in most of this talk).
For theories in which only vector-like symmetries are gauged,
like QCD, this works fine.  The theory
can be made gauge invariant by inserting the SU(3)-color link
variables on each hopping term.  The global chiral symmetry is broken,
but can be restored in the continuum limit by subtracting the quark mass.
However, the situation changes dramatically when we wish to gauge a
chiral symmetry.  We can still try to remove the doublers
with a Wilson mass term, by introducing a
RH ``spectator" fermion $\psi_R$ for each LH fermion $\psi_L$.
(Other possibilities exist, but
the conclusions are similar in all cases \cite{ysnogo}.)  But, now we are
interested in gauging a chiral symmetry, and the Wilson mass term
does not respect gauge invariance (see below).   This means that, on the lattice,
the longitudinal gauge field (which represents the gauge degrees
of freedom ({\it gdofs})) couples to the fermions.  If we only have a
term $\sim{\rm tr}\;F_{\mu\nu}^2$ controlling the dynamics of the gauge
field, the longitudinal modes are not suppressed at all, and their
random nature
typically destroys the chiral nature of the fermion spectrum
(see refs.~\cite{ys,dp} for reviews).
This phenomenon is
non-perturbative in nature:  the problem is invisible for ``smooth"
gauge fields, but the point is that longitudinal gauge fields do
not
have to be smooth, even for small gauge coupling, if all gauge fields
on any orbit have equal weight in the partition function.

This is precisely where gauge fixing comes in.  A renormalizable
choice of gauge adds a term to the gauge-field action which controls
the longitudinal part of the gauge field.  In this talk, we will
consider the Lorentz gauge, with gauge-fixing lagrangian
$(1/2\xi){\rm tr}\;(\partial_\mu A_\mu)^2$ (in its continuum form).
The longitudinal
part of the gauge field ($\partial_\mu A_\mu)$ has now acquired the
same ``status" as the transverse part ($F_{\mu\nu}$), because
the gauge-fixing term acts as a kinetic term for the longitudinal
part of the gauge field, suppressing gauge configurations with
large field components and/or large momenta.

Before we get to the explanation of how this works in detail,
it is instructive to review briefly what goes wrong without
gauge fixing, using our example
of a Wilson mass term.  If we perform a gauge transformation 
$\psi_L\to\phi^\dagger\psi_L$,
$\psi_R\to\psi_R$ (this is 
what it means for $\psi_R$ to be a spectator), with
$\phi$ a group-valued scalar field, the Wilson mass term transforms
into
\begin{equation}
-\frac{r}{2}\sum_\mu\left(\psibar_{Rx}\phi^\dagger_{x+\mu}\psi_{Lx+\mu}
+\dots\right)\;. \label{WY}
\end{equation}
The parameter $r$ is promoted to a Yukawa-like coupling,
and the lattice regulator
leads to couplings between the fermions and 
the longitudinal degrees of freedom represented by 
the scalar field $\phi$.
Note that the lattice theory {\it is} invariant under the symmetry
$\psi_{L,R}\to h_{L,R}\psi_{L,R}$, $\phi\to h_L\phi h^\dagger_R$
{\it etc.}, with $h_R$ global and $h_L$ local ($h$-symmetry).
The $h_L$-symmetry is, however, not
the same as that of the gauge theory we wish to construct, since
$\phi$, representing the longitudinal part of the gauge field,
is supposed to decouple.

We can now explore the phase diagram ({\it i.e.} all values of $r$)
in order to see whether we might decouple these longitudinal modes, while
keeping the fermion spectrum intact.  (In a confining theory,
the ``chiral quarks" do not appear in the spectrum, but we may
first consider the theory with only the
$\phi$
dynamical, with external smooth transverse gauge fields;  the
``reduced model" of refs.~\cite{wmypd,wmyprl,wmylat97}.)
It turns out that three things can happen (see ref.~\cite{dp} and
refs. therein).  First,
$h$-symmetry can be spontaneously broken, and the doublers will
be removed if $\langle\phi\rangle\sim 1/a$.  However, in that case
also the gauge-field mass will be of order $1/a$, which is not what we
want.  It follows that we would like the $h$-symmetry to be unbroken.

For small $r$, we may read off the fermion spectrum by replacing
$\phi\to\langle\phi\rangle$.  If $\langle\phi\rangle=0$, we find
that
the Wilson--Yukawa term does not lead to any fermion masses, and the
doublers are degenerate with the massless physical fermion!  (In
the
broken phase, this degeneracy is partially removed, but, as we already
noticed, the doubler masses will be set by the scale of the gauge-field
mass.  An elegant way of doing this was reviewed in ref.~\cite{montvay}.)
There also exists a phase with unbroken $h$-symmetry at large $r$,
but it turns out that in that phase the only massless LH fermion
is described by the composite field $\phi^\dagger\psi_L$.  This
fermion does not couple to the gauge field, since its
gauge charge is ``screened" by the longitudinal field $\phi$
\cite{mdj,mdrome}.  Both this composite LH and the spectator
RH fermion do not couple to the gauge field, and again,
we failed to construct the desired ChGT, in which
only LH fermions couple to dynamical gauge fields.

What we will show in the rest of this talk is that this conclusion,
that there is no place in the phase diagram where a
ChGT
can be defined, changes completely when  a gauge-fixing term is added,
thus enlarging the parameter space of the phase diagram.

Before we end this introduction,
we would like to rephrase our conclusions thus far in a
somewhat different way.  Imagine that we have defined the fermionic
partition function $Z_F(A)$ for an external lattice gauge field $A$ (not
necessarily smooth!) in a certain
attempt to construct a ChGT.   This then yields
an effective action $S_{\rm effective}(A)=-\log(Z_F(A))$,
and we have, under a gauge transformation,
\begin{equation}
\delta S_{\rm effective}(A)={\rm anomaly}(A)+{\rm lattice\ artifacts}(A)\;.
\label{SEFFECTIVE}
\end{equation}
The anomaly part can be identified by
choosing the external gauge field to be small and smooth (in lattice units).
The lattice-artifact terms are generically {\it not} small.
We know this, because there is no ``small parameter'' to control them,
and also since (as in our simple Wilson-fermion example above)
the dynamics of the longitudinal modes can change the
theory into a vector-like one, that has no gauge anomaly.
In other words, if gauge invariance is
broken by the lattice regulator,  one has to
worry about the {\it back reaction} of the gauge fields {\it on}
the fermion spectrum.
The approach based on the Ginsparg--Wilson
relation constitutes an elaborate algebraic framework
that strives to achieve $\delta S_{\rm effective}(A)=0$ exactly 
on the lattice for
fermion representations which are anomaly-free in the usual,
continuum sense.  In our approach here, the aim is to
control the back reaction through gauge fixing in such a way
that gauge invariance is recovered in the continuum limit,
maintaining the chiral nature of the fermion spectrum
({\it i.e.} without introducing doublers dynamically).

In the next section, we will discuss the gauge-fixing approach
in more detail, concentrating on the abelian case, for which we
claim that our construction is complete.  While we will not address
the generalization to the non-abelian case in general,%
\footnote{For some steps in this direction, see ref.~\cite{bgos}.}
we will, in section 3, discuss how fermion-number violating
processes, such as those known to occur in the Standard Model,
are realized in our approach.

\section{Gauge fixing -- the abelian case}

The central idea of the gauge-fixing approach is to make gauge fixing
part of the definition of the theory \cite{rome}.  This contrasts with the case
of lattice QCD, where, because of the compact nature of the lattice
gauge fields and exact gauge invariance on the lattice,
gauge fixing is not needed.  The theory is defined
by
the action
\begin{equation}
S=S_{\rm gauge}+S_{\rm fermion}+S_{\rm g.f.}+S_{\rm ghosts}+S_{\rm c.t.}\;.
\label{ACTION}
\end{equation}
For $S_{\rm gauge}$ we choose the standard plaquette term.  For
$S_{\rm fermion}$ we use Wilson fermions, with only the LH
fermions coupled to the gauge fields.  We will choose the Wilson
mass
term as in eq.~(\ref{WMASS}), without any gauge fields in the hopping
terms.  Other choices are possible, but in a ChGT, all
break the gauge symmetry.  Our choice has the (technical) advantage of
making the action invariant under shift symmetry, $\psi_R\to\psi_R
+\epsilon_R$, with $\epsilon_R$ a constant, RH Grassmann
spinor \cite{gp}.  For $S_{\rm g.f.}$ we will choose a lattice
discretization of $\int d^4x (1/2\xi)(\partial_\mu A_\mu)^2$, to be
discussed below.  Since here we will only discuss
abelian theories, $S_{\rm ghosts}$ can be omitted \cite{wmynoghosts}.

Since the lattice regulator breaks gauge invariance explicitly, counter terms
are needed, and they are added through $S_{\rm c.t.}$.  These counter terms
include one dimension 2 operator (the gauge-field mass counter term),
no dimension 3 operators (because of the shift symmetry), and a host of
dimension-4 counter
terms (see refs.~\cite{rome,wmky} for a detailed discussion).
Tuning these counter
terms to the appropriate values (by requiring the Ward--Takahashi
identities of the continuum target theory to be
satisfied) should then bring us
to the critical point(s) in the phase diagram at which a
ChGT can be defined.  Because of the choice of a renormalizable gauge,
it is clear that this can be done in perturbation theory (if the
theory
is anomaly free).  The observation of ref.~\cite{rome} is that also
non-perturbatively gauge-fixing will be needed in order to make this
all work.

At the non-perturbative level, the following important questions arise
\cite{ysrome}.  First, what should we choose as the lattice discretization
of $S_{\rm g.f.}$?  More precisely, given a certain choice, what
does
the phase diagram look like, and for which choices do we find a phase
diagram with the desired critical behavior?  Second, if we find that a
suitable discretization exists, so that the fermion content is indeed
chiral, how does this precisely happen?  Third, since fermion-number
violating processes typically occur in ChGTs, how
does our method provide for them?  Note that, without gauge fixing,
the action above is essentially just the Smit--Swift model \cite{ssm},
which, as summarized in the introduction, does not work.
In this section, we will sketch the answers
to the first two questions.  We postpone fermion-number violation
to the next section.  The complete extension of our ideas to the
non-abelian case requires a better understanding of the
non-perturbative aspects of gauge fixing, and we will not discuss
this in this talk.

\subsubsection*{Gauge fixing on the lattice}
\medskip
It was argued in ref.~\cite{ysrome} that the lattice gauge-fixing term
\begin{equation}
S_{\rm g.f., naive}=\tk\sum_x\left(
\sum_\mu({\rm Im}\;U_{x,\mu}-{\rm Im}\;U_{x-\mu,\mu})\right)^2 \,, \qquad
\tk=\frac{1}{2\xi g^2} \,,
\label{NAIVE}
\end{equation}
is {\it not} the right choice, even though, expanding the link
variables $U_{x,\mu}={\rm exp}(igA_{x,\mu})$, it looks like a
straightforward discretization of the continuum form.  This is because
this choice admits an infinite set of lattice Gribov copies (which
have no continuum counter part) of the perturbative vacuum $U_{x,\mu}=1$.
This is dangerous because lattice Gribov copies mean
large longitudinal modes which, as we have explained,
can spoil the fermion spectrum.
Therefore, we insist that {\it lattice perturbation theory should be a
reliable approximation of our
lattice theory at weak coupling}.  In fact, we showed, through a
combination of numerical and mean-field techniques, that the naive
choice of gauge-fixing action of eq.~(\ref{NAIVE}) does not lead
to
a phase diagram with the desired properties \cite{wmky}.

The vacuum degeneracy of $S_{\rm g.f., naive}$ can be lifted by adding
irrelevant terms to it \cite{ysrome,my}, so that
\begin{equation}
S_{\rm g.f.}=S_{\rm g.f., naive}+{\tr}\tk\, S_{\rm irrelevant}\;,
\label{GAUGEFIX}
\end{equation}
where $\tr > 0$ is a parameter very similar to the Wilson parameter
$r$ multiplying the Wilson mass term.  While we will not give any
explicit form of $S_{\rm irrelevant}$ here, it was shown \cite{my}
that $S_{\rm irrelevant}$ can be chosen such that
\begin{equation}
S_{\rm g.f.}(U)\ge 0\ \ \ {\rm and}\ \ \ S_{\rm g.f.}(U)=0
\Leftrightarrow U_{x,\mu}=1\;.\label{PROP}
\end{equation}
This means that $U_{x,\mu}=1$ is the unique perturbative vacuum.
Also, obviously, we still have
$S_{\rm g.f.}(U)\to \int d^4x (1/2\xi)(\partial_\mu
A_\mu)^2$ in the classical continuum limit.  Our choice does not
respect BRST symmetry, so that we will need to adjust
counter terms \cite{wmynoghosts}.

For small gauge coupling $g$, the classical potential should give us
an idea of what the phase diagram looks like.  Without fermions
(which contribute to the gauge-field effective potential only at
higher orders in lattice perturbation theory), including (only) a
mass counter term $-\kappa\sum_\mu(U_{x,\mu}+U^\dagger_{x,\mu})$,
and
expanding $U_{x,\mu}={\rm exp}(igA_{x,\mu})$, we have, for our choice of
$S_{\rm irrelevant}$,
\begin{equation}
V_{\rm classical}(A)=\frac{{\tr}\tk g^6}{2}\left(\sum_\mu A^2_\mu
\sum_\nu A^4_\nu+\dots\right)+\kappa g^2\left(\sum_\mu A^2_\mu+\dots
\right)
\label{CLASS}
\end{equation}
for a constant field.  The dots indicate higher-order terms in $g^2$.
While the precise form of the term proportional to $\tr$ is not
important, it is clearly irrelevant and positive
({\it i.e.} it stabilizes the perturbative vacuum).

We can now distinguish two different phases, depending on the value
of $\kappa$.  For $\kappa>0$,
$\langle A_\mu \rangle=0$, and the gauge field has a
positive mass $\sqrt{2\kappa g^2}$.  For $\kappa<0$, the gauge field
acquires an expectation value
$\langle gA_\mu \rangle=\pm\left(-\frac{\kappa}
{6\tr\tk}\right)^{1/4}$, for all $\mu$, and we encounter a
new phase, in which the (hyper-cubic) rotational symmetry is
spontaneously broken!
These two phases are separated by a continuous phase transition
(classically at $\kappa=\kappa_c=0$), at which the gauge-field mass
vanishes.  It follows that we are interested in taking the
continuum limit by
tuning $\kappa\searrow\kappa_c$.  (For a discussion including
all dimension-four counter terms, see ref.~\cite{my}.)

A detailed analysis of the phase diagram for the abelian
theory without fermions was given in ref.~\cite{wmky}.  A complete
description of the phase diagram in the four-parameter space
spanned by the couplings $g$, $\tk$, $\tr$ and $\kappa$ can
be
found there, as well as a discussion of the other counter terms and
a study of gauge-field propagators.
In the region of interest (small $g$, large $\tk$ and $\tr\approx 1$)
good agreement was found between a high-statistics numerical study and
lattice perturbation theory.  The picture that emerges
from the classical potential as described above was shown to be
correct, as long as we choose $\tr>0$ away from zero,
and the coupling constants $g^2$ and $\tk^{-1}=2\xi g^2$ sufficiently small.
As it should,
the theory (without fermions) at the critical point describes (free)
relativistic photons.

\subsubsection*{Fermions}
\medskip
We now come to the behavior of the fermions in this gauge-fixed lattice
theory.  Employing a continuum-like notation for simplicity, our
lattice
lagrangian, including fermions, reads
\begin{eqnarray}
{\cal L}&=&\frac{1}{4}F_{\mu\nu}^2+\tk g^2(\partial_\mu A_\mu)^2
+{\tr \tk}{\cal L}_{\rm irrelevant}(gA) \label{VECTOR} \\
&&+\psibar\left(\Dsl(A)P_L+\dsl P_R\right)\psi-\frac{r}{2}\psibar\Box\psi
\nonumber \\
&&+\kappa g^2 A_\mu^2+\mbox{other counter terms}\;. \nonumber
\end{eqnarray}
In order to investigate the interaction between fermions and longitudinal
modes, we can make the latter explicit by a gauge transformation
\begin{equation}
A_\mu\to\phi^\dagger A_\mu\phi-\frac{i}{g}\phi^\dagger\partial_\mu\phi
\equiv -\frac{i}{g}\phi^\dagger D_\mu\phi\;,
\ \ \ \  \psi_L\to\phi^\dagger\psi_L\;.\label{GAUGETR}
\end{equation}
This yields the lagrangian in the ``Higgs" or ``St\"uckelberg" picture,
\begin{eqnarray}
{\cal L}&=&\frac{1}{4}F_{\mu\nu}^2+\tk\left(\partial_\mu
(\phi^\dagger(-i\partial_\mu+gA_\mu)\phi)\right)^2
+{\tr \tk}{\cal L}_{\rm irrelevant}(gA,\phi) \label{HIGGS} \\
&&+\psibar\left(\Dsl(A)P_L+\dsl P_R\right)\psi-\frac{r}{2}\left(\psibar_R
\Box(\phi^\dagger\psi_L)+\psibar_L\phi\Box\psi_R\right)
\nonumber \\
&&+\kappa\left(D_\mu(A)\phi\right)^\dagger
\left(D_\mu(A)\phi\right)+\mbox{other counter terms}\;, \nonumber
\end{eqnarray}
which is invariant under the $h$-symmetry mentioned in the
introduction.

In order to find out whether the longitudinal modes, which are
represented by the field $\phi$ in the Higgs-picture lagrangian,
change the fermion spectrum, we may simplify the theory by
considering the ``reduced" model, in which we set $A_\mu=0$
in eq.~(\ref{HIGGS}).  Expanding $\phi={\rm exp}(i\theta/\sqrt{2\tk})$,
which is appropriate for small $g$ because $1/\sqrt{\tk} \propto g$
(see eq.~(\ref{NAIVE})), gives the reduced-model lagrangian
\begin{eqnarray}
{\cal L}_{\rm red}
&=&\frac{1}{2}(\Box\theta)^2+\frac{\kappa}{2\tk}(\partial_\mu\theta)^2
+\mbox{irrelevant}\ \theta\ \mbox{self-interactions} \label{REDUCED} \\
&&+\psibar\dsl\psi-\frac{r}{2}\psi\Box\psi
+\frac{i}{\sqrt{2\tk}}
(\psibar_L\theta\Box\psi_R-\psibar_R\Box(\theta\psi_L))+
O(g^2)\;. \nonumber
\end{eqnarray}
This lagrangian teaches us the following.  First, $\theta$ is
a real scalar field with dimension 0, and inverse propagator
$p^2(p^2+\kappa/\tk)$.  Near the critical point (which is
at $\kappa_c=0$ to lowest order), this behaves like $p^4$.  This
actually implies \cite{wmypd,wmypt} that
\begin{equation}
\langle\phi\rangle\propto (\kappa-\kappa_c)^{1/(32\pi^2\tk)}\to 0
\label{VEV}
\end{equation}
for $\kappa\to\kappa_c$.  (This behavior is very similar to that
of
a normal scalar field in two dimensions in the massless limit.)
This means that $h$-symmetry, which is spontaneously broken
on the lattice, gets restored {\it at} the critical
point.

The fermion-scalar interactions in eq.~(\ref{REDUCED}) are
dimension 5, and therefore irrelevant.  This (heuristically)
implies that $\theta$, which represents the longitudinal modes or {\it gdofs},
decouples from the fermions near the critical
point.  The doublers are removed by the Wilson mass term,
which is present in eq.~(\ref{REDUCED}).  The conclusion is that
a continuum limit exists (at the critical point of the reduced model)
with free charged
LH fermions ({\it i.e.} fermions which couple to the
transverse gauge field in the full theory)
and free neutral RH fermions (the spectators).
In other words, the fermion spectrum is chiral.  It is clear from
the discussion here that gauge-fixing plays a crucial role: without
it, the higher-derivative kinetic term for $\theta$ would not be
present.  It is the infrared behavior of $\theta$ that
causes this novel type of critical behavior to occur.  Note,
finally, that the restoration of $h$-symmetry at the critical point
and the decoupling of $\theta$ from the fermion fields together imply
that the target gauge group is unbroken in the resulting continuum
theory.

Of course, the description given here is quick and dirty.  The unusual
infrared properties of this theory were investigated perturbatively
in much more detail in ref.~\cite{wmypt}.  Fermion propagators
were computed numerically in ref.~\cite{wmyprl}, and the agreement
with perturbation theory was shown to be very good.
(The numerical computations were done in
the quenched approximation.  However, the effects of quenching
occur only at higher orders in perturbation theory, so the
good agreement between numerical and perturbative results
indicates that this is not a serious problem.)  All these
studies confirm the results described in this talk.

\section{Fermion-number violation}

In this section, we will briefly describe how fermion-number
violating processes occur in our approach
approach \cite{fnv}.
It was observed a while ago that this is
a non-trivial issue \cite{banks}.  The problem originates
in the fact that, in the gauge-fixing approach, flavor symmetries
which are anomalous in the continuum appear to be
conserved on the lattice.

It is easiest to explain the issue by considering one-flavor QCD.
Consider the following lagrangian for a LH quark $\psi_L$ with
a RH spectator $\chi_R$ and a RH quark $\psi_R$ with a LH
spectator $\chi_L$:
\begin{eqnarray}
\label{ONEFLAVOR}
{\cal L}&=&\psibar_L\Dsl\psi_L+\chibar_R\dsl\chi_R
-\frac{r}{2}(\chibar_R\Box\psi_L+{\rm h.c.})\\
&&+\psibar_R\Dsl\psi_R+\chibar_L\dsl\chi_L
-\frac{r}{2}(\chibar_L\Box\psi_R+{\rm h.c.})\,.\nonumber
\end{eqnarray}
Of course, this is not the simplest way of putting one-flavor QCD
on the lattice, but it helps illuminating the problem and its
solution in a familiar context.  For $r=0$, this theory has
four conserved fermion-number symmetries, broken to two $U(1)$
symmetries when $r\ne 0$:
\begin{equation}
\label{UONES}
U(1)^\chi_R\times U(1)^\psi_L\times
U(1)^\psi_R\times U(1)^\chi_L
\to U(1)^\psi_L\times U(1)^\psi_R\cong U(1)_V\times U(1)_A\,,
\end{equation}
{\it i.e.} the theory has ``too much" symmetry, since $U(1)_A$
should be anomalous!

In perturbation theory, the resolution is rather straightforward
\cite{dugman,wmypt}.
The corresponding axial current ${\hat J}_\mu^A$
is exactly conserved on the lattice, but not gauge invariant.  One
may define a new current $J_\mu^A={\hat J}_\mu^A+g^2 K_\mu$
with $K_\mu$ a lattice operator with continuum limit
\begin{equation}
\label{KMU}
K_\mu=\frac{1}{16\pi^2}\epsilon_{\mu\nu\rho\sigma}
{\rm tr}\;(A_\nu F_{\rho\sigma}-\frac{1}{3}A_\nu A_\rho
A_\sigma)\,.
\end{equation}
It can then be shown that $J_\mu^A$ is gauge invariant, and has
a divergence equal to the usual anomaly.

However, if we go beyond perturbation theory, this does not solve
the basic question how, in a one-instanton background,
$\psibar_R\psi_L$ can pick up a non-vanishing expectation value
\cite{banks}.  (The presence of the spectator field $\chi$
does not help: there is no term in the lagrangian connecting
$\psi_L$ and $\psi_R$.)  The only way out is to demonstrate that
{\it spontaneous} symmetry breaking (SSB) occurs \cite{kosu,coleman}.
The procedure to follow then is the following.
\begin{itemize}
\item
Turn on the appropriate infinitesimal external ``magnetic" field,
in this case a small quark mass, $m\psibar\psi$;
\item
Take the volume to infinity, the lattice spacing to zero, and
finally $m\to 0$ (relative to physical scales);
\item
See if $\langle\psibar_R\psi_L\rangle\ne 0$ in an instanton
background for $m\to 0$;
\item
In order to use semi-classical methods, take the instanton
size $\rho\gg a$, but small enough that the renormalized
coupling constant is small.
\end{itemize}
The Dirac operator ${\cal D}$ corresponding to eq.~(\ref{ONEFLAVOR})
is
\begin{equation}
\label{DIRAC}
{\cal D}\pmatrix{\chi_R\cr\psi_L\cr\psi_R\cr\chi_L}
=\pmatrix{0&0&-r\Box/2&\overline\sigma_\mu\partial_\mu\cr
0&m&\sigma_\mu D_\mu&-r\Box/2\cr
-r\Box/2&\overline\sigma_\mu D_\mu&m&0\cr
\sigma_\mu\partial_\mu&-r\Box/2&0&0}
\pmatrix{\chi_R\cr\psi_L\cr\psi_R\cr\chi_L}\,.
\end{equation}
Consider now the LH zero mode $\Psi=\pmatrix{0\cr u_L\cr 0\cr 0}$
with $\Dsl u=0$, $u=\pmatrix{u_L\cr 0}$, in the continuum.
Latticizing this zero mode, one has that
\begin{equation}
\label{ZM}
{\cal D}\Psi=(m+{\cal O}(a/\rho^2))\Psi\,.
\end{equation}
For $a\to 0$, $\chi$ decouples, ${\cal D}\Psi\to(\Dsl+m)u=mu$, and
one finds that (for technical details, see ref.~\cite{fnv})
\begin{eqnarray}
\label{ZMPROP}
\lim_{a\to 0}{\cal D}^{-1}(x,y)
&=&\frac{1}{m}u(x)u^\dagger(y)+{\cal O}(1)\,,\\
\lim_{a\to 0}{\rm det}\;{\cal D}&=&
m({\rm det}'\;+{\cal O}(m))\,,\nonumber
\end{eqnarray}
for small $m$, where ${\rm det}'$ is the determinant with the
zero mode removed.  Putting things together, we find the
desired 't Hooft vertex \cite{thooft}
\begin{equation}
\label{FINAL}
\lim_{m\to 0}\langle\psibar_R(x)\psi_L(y)\rangle
=
u^\dagger(x)u(y)\;{\rm det}'\,.
\end{equation}
It can be shown that, even though the 't Hooft vertex appears
through SSB, there are no gauge-invariant Goldstone poles
\cite{fnv,kosu,coleman} in the continuum limit.
Essentially, the explanation is that
${\hat J}_\mu^A$ is not gauge invariant.

In order to see whether this mechanism also works in a genuinely
ChGT, we also worked out the example of an SO(10)
theory with a LH Weyl fermion in the 16-dimensional representation
of SO(10) \cite{fnv}.  In this case, there are four
independent LH zero modes in a one-instanton background 
(one can embed SU(2)$^4$ in SO(10)).  The
symmetry breaking mass term is chosen to be
\begin{equation}
m\left(\psi_L^T\epsilon C\psi_L+\psibar_L\epsilon C\psibar^T_L
\right)\,, \label{TENDMASS}
\end{equation}
where $\epsilon$ acts on the Weyl index,
and $C$ is a symmetric charge-conjugation-like matrix
acting on the SO(10) index.  This mass term breaks the U(1)
fermion-number symmetry, which is anomalous in the continuum
target theory.  The fermion determinant in the instanton
background is proportional to $m^2$ (in the limit of
vanishing lattice spacing), and one can show that, as a
consequence, one obtains the expected four-fermion 't Hooft
vertex
\begin{equation}
\lim_{m\to 0}\langle\psi_L(x)\psi_L(y)\psi_L(z)\psi_L(w)\rangle
=\epsilon_{ijkl}u_i(x)u_j(y)u_k(z)u_l(w)\;{\rm det}'\,,
\label{FFVERTEX}
\end{equation}
where $i,j,k,l=1,2,3,4$ label the four different zero modes.
We see that again 't Hooft vertices arise through SSB of the
lattice fermion-number U(1) symmetry.  We note that the SO(10)
theory is rich enough to
contain the Standard Model, as well as many interesting
Grand Unified Theories.

\section{Conclusion}

Let us summarize the progress reviewed in this talk.  We have
demonstrated how gauge fixing on the lattice can be used to solve
the problem of coupling lattice fermions chirally to gauge fields.
The method works for abelian theories, where no ghosts (or
anything equivalent) are needed.  Whether we can complete this
proposal for constructing lattice ChGTs also
for non-abelian theory depends solely on whether the
non-perturbative gauge-fixing problem can be solved for this case.
An attractive feature is the fact that this method can in principle
be applied to any lattice fermion method, thus showing a degree
of universality.  For an investigation using domain-wall fermions,
see ref.~\cite{bd}.  New support for this method comes from the
fact that there are no surprises with respect to fermion-number
violating processes; things work basically just as one would
expect in the continuum when one would employ a regulator that
breaks gauge invariance.  (For earlier work on the interplay
between gauge invariance and fermion-number violation, see
ref.~\cite{bhs}.)

\subsubsection*{Acknowledgements}
\medskip
We would like to thank Pierre van Baal and Aharon Casher
for useful discussions, and Wolfgang Bock and Ka Chun Leung for
collaboration on much of this work.   MG would like to thank the
organizers of the workshop ``Confinement, Topology, and other
Non-Perturbative Aspects of QCD," where this talk was given,
for a very pleasant and well-organized conference.
Both of us thank the Institute for Nuclear Theory at
the University of Washington for hospitality and support provided
for part of this work.
This research is supported in part by a grant from the
United-States -- Israel
Binational Science Foundation, and by the US Department of Energy.

\end{document}